Title of the paper: **Narrowband IoT for Healthcare**

Name of the author(s): **Dr Sudhir K. Routray and Sharath Anand**

Name of the college & University: **CMR Institute of Technology, Bangalore**

Complete postal address with email id and telephone numbers:
    **Department of Telecommunication Engineering**
    **CMR Institute of Technology, Bangalore**
    **#132 AECS Layout, ITPL Main Road, Kundalahalli**
    **Bangalore - 560037**
    **Email: sudhirkumar.r@cmrit.ac.in / shan15dc@cmrit.ac.in**
    **Phone: +91 8028524466**

# Narrowband IoT for Healthcare

Sudhir K. Routray
Associate Professor, Telecommunication Engineering
CMR Institute of Technology, Bangalore
India
sudhirkumar.r@cmrit.ac.in

Sharath Anand
Telecommunication Engineering
CMR Institute of Technology, Bangalore
India
shan15dc@cmrit.ac.in

*Abstract*—The Internet of Things (IoT) is going to have its presence in all the essential sectors of human lives. It has the ability to provide both mainstream as well as the value added services in almost all the sectors. Healthcare is an important service sector for overall development. It has far reaching implications in the quality of living. In the modern world where the quality of living has been degraded significantly IoT can certainly play a constructive role in providing better services. In healthcare, there are several occasions such as patient health monitoring, remote observation and emergency proceedings outside the hospital where sensors can play essential roles. The coordinated sensor networks can provide even better services. IoT has the ability to provide all these coordinated services. Narrowband IoT (NBIoT) is an economical and simpler version of IoT which can handle these tasks effectively. Due to the widespread requirement of healthcare, NBIoT is a preferred solution as it needs fewer amounts of resources. In this article, we provide the main issues and difficulties of NBIoT in healthcare.

## I. INTRODUCTION

The Internet of Things (IoT) is a sensor based technology that can connect several objects to the Internet. This connectivity can provide us several services which are essential for the day to day activities. Healthcare sector is very much important and several sensor based applications are required in its proper decision making and pre-treatment proceedings. In addition to that, there are several other instances where the remote monitoring of patients is essential for successful treatment. The resources required for wide spread healthcare services is large. Thus a resource efficient IoT is required for these applications. Narrowband IoT (NBIoT) is certainly one of the resource efficient forms of currently available IoTs. We propose, NBIoT as a potential solution for the healthcare related applications.

IoT based solutions for healthcare have been proposed in several works. In [1], a comprehensive survey of healthcare IoT has been presented. It provides all the main aspects of current initiatives for IoT based healthcare. It also provides several potential architectures of IoT for effective healthcare service provisioning. In [2], an energy efficient and trustworthy healthcare IoT system is proposed which can be remotely monitored. In this work, the authors have emphasized on the security and energy consumption aspects of IoT based healthcare system. In [3], a body sensor network has been proposed for monitoring several health parameters. Security aspects have also been considered in this work which proposes robust security measures for healthcare IoT. A similar approach has been proposed in [4] which take care of the health monitoring using a group of connected sensors. It also uses body sensor networks for this purpose. In [5], an IoT aware healthcare system has been proposed for smart healthcare systems. In this case, the IoT sensors take measurement of various health related parameters and send them to a central hub where the information is processed and decisions are taken for further course of action. In [6], a low power sensor design methodology has been proposed for mobile based healthcare applications. Wide area coverage with low power is the main aim of this IoT based work. Radio frequency identification (RFID) based personal healthcare in the smart homes and other smart spaces has been proposed in [7]. It uses IoT based connected architecture for personal health motoring. NBIoT is very popular in the low power applications. In [8], the main standards issues of NBIoT and its potential applications have been presented. Main attraction of NBIoT is that it has already been standardized in LTE Release 13. In [9], the main points of LTE Release 13 for NBIoT are documented. It provides the physical details and deployment options for NBIoT. In [10], the main applications of NBIoT are presented in which healthcare is one such potential areas. Modern healthcare systems have several different complexities and NBIoT is found to be suitable for this complex scenarios.

In this article, we present the suitability of NBIoT for healthcare applications. We provide the details of NBIoT, its physical parameters, the basic standardization issues as presented in Release 13, and the matching characteristics with the healthcare needs. We also analyze the potential demerits of this technology and suggest the alternatives to overcome them.

The remaining parts of this paper are organized in five different sections. In Section II, we present the basic features of NBIoT and its potential applications. In Section III, we present the utilities of IoT for healthcare. In Section IV, we present the justification and practical utilities of NBIoT for healthcare. In Section V, we present the potential difficulties and demerits of NBIoT for this sector. In Section VI, we conclude this article with the main points.

## II. NARROWBAND IOT

NBIoT is a leaner and thinner version of IoT. It takes a narroband of frequency for its operation. In Release 13, 180 kHz is allocated for NBIoT [8]. It can be deployed in both cellular and noncellular forms. Noncellular forms are needed in some ad hoc applications. Cellular forms are popular as they are very organised and can use the cellular infrastructure for their operations.

In Realease 13, cellular IoT for GSM and LTE networks are standardised in terms of the operating parameters. Three different types of depolyment has been proposed for NBIoT [9]. The first one is the standalone deployment in which a new band of microwave frequencies have been proposed. Right now, the microwave frequencies in the 700 MHz and 800 MHz are popular for NBIoT standalone deployments. In the second option the guard bands of LTE and GSM are proposed for NBIoT operations. These guard bands are notused by LTE and GSM operators. So, these bands can be used for NBIoT for value added services in the cellular networks. The third option is the in-band deployment of NBIoT. In this case, some of the opeational bands of LTE and GSM is provided for NBIoT operations. For this to happen in a systematic way, appropriate frequency hopping algorithms are provided. NBIoT will take several legacy propoerties form GSM and LTE. Therefore, the layerwise architecture of NBIoT will have several similar features of GSM and LTE.

NBIoT is the main attraction in several applications because it is a low power wide area (LPWA) coverage capabilities. For the widespread application sectors such as the healthcare LPWA technologies are essential. NBIoT has two different input power specifications: 20 dBm and 23dBm. It has twotier power saving mechanisms in place [9]. It uses half duplex communication for its operations. Binary phase shift keying (BPSK) and quadrature phase shift keying (QPSK) are the two main modulation schemes used in NBIoT at the moment. Therefore, the peak downlink and uplink dat rates possible right now is around 250 kbps. It can cover a power strength as low as -164 dBm. This is what shows NBIoT is a real LPWA technology.

## III. IOT FOR HEALTHCARE

Main motivations for IoT in healthcare come from two obvious reasons. The first is the pre-treatment procedures required for an emergency when the patient is being brought to the hospital. The availability of direct contact with the patient and remote monitoring by the healthcare experts along the way to the hospital can save many lives. The support staff can take appropriate decisions and provide essential services when they are advised by the experts. The second reason is that several ailments need constant monitoring. It is not possible to keep these people in the hospital as they do not need the medical services all the time. Rather only during the onset of the problems they need to see the doctors. In such cases, IoT is very much desirable. It sends the health information to the healthcare service provider regularly and when there is an urgent need of intervention of the experts. On the other end, the healthcare provider keeps monitoring the information received from the IoT and takes the appropriate steps as they are required.

In addition to the above, IoT has several advantages of providing remote monitoring is several applications including healthcare [1]. Remote monitoring of health is required in several occasions in which the real-time information tracking is advisable. As the IoT is going to be an intelligent service provider and body area sensors are readily available IoT for healthcare is certainly an attractive technology [3]. Of course providing healthcare from a remote location has several challenges. The information being transmitted in a wireless channel has several potential threats [2]. These issues have to be handled appropriately for safe and secure healthcare.

## IV. NBIoT FOR HEALTHCARE

As we have seen in Section II, NBIoT is the attraction in sectors which are widespread as it is a LPWA technology. Certainly, healthcare is one such sector and it needs this technology [10]. Connectivity among sensors paved the way for gathering important information that has not been possible in the past. Advancements have been made in the sensor networks and communication with the evolved technologies, and thereby paving way for it to be adapted to various real time fields. Healthcare is one such field where we can use NBIoT as an alternative and as an easy technology for diagnosing the variations in the functioning of human body. Blood pressure, heart rate, vital capacity and many other parameters along with their functions can be recorded and analyzed by installing the appropriate sensor. This allows us to collect the patient data at regular intervals and preventive measures can be taken. In addition to the above, NBIoT has the compatibility with the existing cellular systems which is an extra advantage for ubiquitous healthcare monitoring.

There are two main scenarios that have to be taken into account, namely clinical care and remote monitoring. In clinical care, the patients are admitted in hospital and their physiological status has to be dealt with vital care. This can be established with NBIoT noninvasive monitoring. This technique employs the collection of physiological data with the help of sensors and the data is stored in local gateways and clouds. The stored data can then be retrieved and passed over to the care takers. This will help to reduce the necessity of periodic health checking of the patient by the health professional. Proper medication can be provided on diagnosing the vital signs and hence proper care can be provided to the patient. This technology helps to reduce the cost and at the same time improves the quality of services provided.

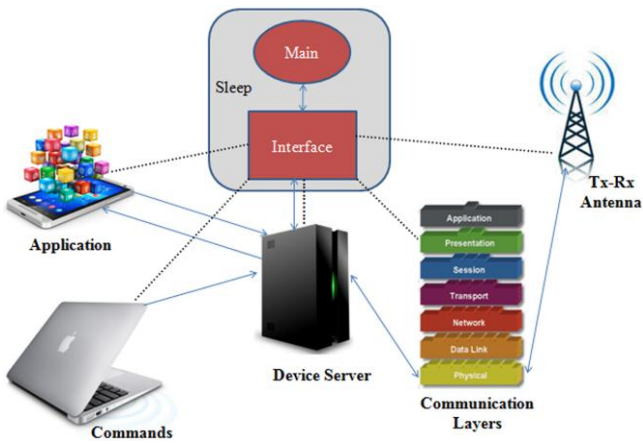

Fig. 1. The basic structure of the device or client side NBIoT for healthcare service provisioning.

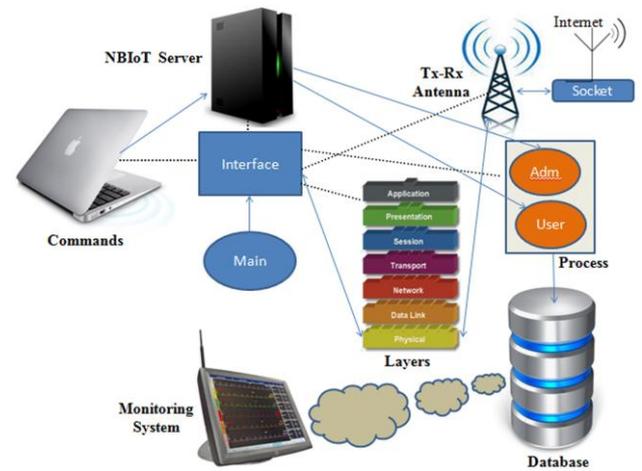

Fig. 2. The basic structure of the server side NBIoT for healthcare service provisioning.

In remote monitoring, the patient is not necessarily admitted in the hospital, but he or she is constantly under the surveillance of the medical device which records the vital signs. The NBIoT device is attached to the human body and it continuously records the different parameters. The recorded information is then transmitted using the transmit antenna present in the device to that of the server present in the caretakers place, where the data will be continuously monitored on a monitor. Hence they can keep a constant eye on the patient and in case of variations, preventive measures can be provided. The client side architecture for NBIoT has been shown in Fig. 1.

The small NBIoT device is attached to that of the patient's body so that wherever the patient may be, the doctors can have complete idea about the patient's health cycle. The implementation of the entire process had two sections: the device structure, and the server structure. The main function invokes the interface function. The interface then creates the server program. Through the keyboard, several physical and medium access control (MAC) layer functions are invoked. Inputs for the configuration is provided through the keyboard, and the configuration messages are directly sent to the device server. An application (App) is used to perform the specific healthcare related actions (for instance, periodic monitoring of the vital signs of the patient). These queues are initiated by the interfaces and after initiating the sub-processes along with the main it goes to sleep mode. The trans-received messages are sent through the internet from the respective ends.

Server in comparison with the device structure has main and interfaces which normally go to sleep mode after the initiation of the sub-process by the interface through the queues which are present. Interface is also responsible for creating the NBIoT server which also has got keyboard through which the configuration messages can be sent. The information present here is routed through the MAC layer and then it is sent to the trans-receiver antenna. The messages then travel through the socket and enter the communication medium and become a part of the 'Internet'. In addition, we have database which can be accessed by processes to store the information. The database can be used for devices as well as users. Users can be nurses, doctors, paramedics and other authorized healthcare officials. The information stored after diagnosing the patient can be viewed from the available visual platform by one or more authorized personnel. The server side architecture for NBIoT has been shown in Fig. 2.

Smart sensors consist of sensors and microcontrollers which are capable of measuring the vital sign of human body accurately. Using these sensors, we can assure that quality service is being provided. This device can also be upgraded according to varying specifications. The microcontrollers can also be reprogrammed so that it is adaptable to the evolving environment. These are also user friendly and cost effective. With the help of these sensors, it is also possible to ensure the amount of intake of medicine and the time of intake. In addition to this, the doctors can ensure that regular medication is in process through the sensor mounted on the medicine bottle. NBIoT sensors require very small amount of power for their operation and the battery lifetime is estimated to be around 13-15 years.

V. POTENTIAL DEMERITS OF NBIOT BASED HEALTHCARE SERVICES

In previous sections, we have already seen the main features and advantages of using NBIoT for healthcare. It has a few demerits as well. The present services provided over the NBIoT platforms are not suitable for real-time applications. The main reason behind this is the lack of control over the delay. The delay tolerant methodologies have not been incorporated in the NBIoT framework as of now. In contrast to this, there are several applications in healthcare that demands real-time communication and information exchange.

In such cases, certainly, we cannot rely on NBIoT. In addition to this, some of the healthcare applications need higher bandwidth for better communication and sensing. In such cases, the bandwidth required is more than what has been allocated to the NBIoT now. With 180 kHz these services will not be achieved as desired with high definition images and better quality of information being exchanged between the clients and the server. The quality of service and quality of experience are two important parameters in IoT related services. At both these fronts NBIoT is not yet at par with the best ICT services available today. Of course, the NBIoT standards will be changed gradually to be compatible with these new demands across several application areas.

Several other possible demerits are also there. Most of the proposed NBIoT solutions for healthcare come as wearable electronics. These wearable devices are not always safe and there are also senses of uncomfortable aspects to these devices. Different manufacturers are having different standards as there is no unison yet for unified methods for manufacturing of IoT devices. It complicates the scenario as the devices used by the hospital and the patients may not match perfectly. Doctors have to be trained according to the IoT practices as they are not aware of these technologies. It would cost a lot to train all the medical staff. Though NBIoT is an economical option for the widespread deployment in the healthcare sector several challenges are still there.

## VI. CONCLUSIONS

NBIoT is one of the leaner, thinner and greener versions of IoT, and it has great potentials in several sectors. It can provide cost effective services which are essential for both rural and urban areas. Healthcare is a primary service sector and it needs smart technologies for better services. In this article, we have shown the main utilities of NBIoT for healthcare. We have gone through the specific functions where it can play important roles in effective service provisioning. We have also indicated the potential demerits due to the current deficiencies in the technology standards. However, these short comings will be overcome in the newer versions of NBIoT. It is expected that NBIoT will be an integral part of modern healthcare system.